\documentclass[journal]{IEEEtran}
\usepackage{cite}
\usepackage[pdftex]{graphicx}
\usepackage{amssymb,amsmath}
\usepackage{float}
\usepackage{bm}
\usepackage{mathtools}
\usepackage{amsmath}
\usepackage{array}
\usepackage{multirow}
\usepackage{stfloats}
\usepackage[utf8]{inputenc}
\usepackage[english]{babel}
\usepackage{array}
\usepackage{mathrsfs}
\usepackage{makecell}
\usepackage{tabularx,booktabs}
\usepackage{hyperref}
\usepackage[most]{tcolorbox}
\usepackage{color, colortbl}
\usepackage{pifont}
\usepackage{amsthm}
\usepackage{midfloat}

\usepackage{algorithm,algpseudocode}
\makeatletter
\renewcommand{\Function}[2]{%
  \csname ALG@cmd@\ALG@L @Function\endcsname{#1}{#2}%
  \def\jayden@currentfunction{#1}%
}
\newcommand{\funclabel}[1]{%
  \@bsphack
  \protected@write\@auxout{}{%
    \string\newlabel{#1}{{\jayden@currentfunction}{\thepage}}%
  }%
  \@esphack
}
\let\myorg@bibitem\bibitem
\def\bibitem#1#2\par{%
  \@ifundefined{bibitem@#1}{%
    \myorg@bibitem{#1}#2\par
  }{%
    \begingroup
      \color{\csname bibitem@#1\endcsname}%
      \myorg@bibitem{#1}#2\par
    \endgroup
  }%
}

\makeatother

\begin{document}
\title{Transient Stability Assessment for Current Constrained and Unconstrained Fault Ride-Through in Virtual Oscillator Controlled Converters}

\author{M A~Awal,~\IEEEmembership{Student Member,~IEEE,}
        and~Iqbal~Husain,~\IEEEmembership{Fellow,~IEEE}
\thanks{This work has been supported in part by the National Science Foundation under award number EEC-0812121 for the FREEDM Engineering Research Center.}
}
\markboth{Awal et. al.: Transient Stability Assessment for Virtual Oscillator Controlled Converters}
{}
\maketitle

\begin{abstract}
Unified virtual oscillator controller (uVOC) inherits the rigorous analytical foundation offered by oscillator based grid-forming (GFM) controllers and enables fast over-current limiting and fault ride-through (FRT). Control design for effective FRT requires transient stability analysis. Existing transient stability analysis methods and studies are limited in either considering only current unconstrained scenarios or neglecting the simultaneous power angle and voltage dynamics. Under current-constrained faults, the voltage and power angle dynamics are strongly coupled and both play critical roles in determining transient stability. Therefore, decoupled analysis of the two, typically used in transient stability studies, does not offer comprehensive insight into the system dynamics. In this work, the overall FRT method for uVOC is presented and a comprehensive modeling and analysis method for transient stability is developed under both current-saturated and unsaturated symmetrical AC faults. We utilize phase-plane analysis of the overall system in a single graphical representation to obtain holistic insights into the coupled voltage and power angle dynamics. The FRT controller and the analysis method have been validated through simulations and hardware experiments. The results demonstrate that uVOC is not constrained by a critical clearing angle unlike droop and virtual synchronous machine (VSM) type second order controllers.
\end{abstract}

\vspace{10pt}

\begin{IEEEkeywords}
Transient stability, unified virtual oscillator control (uVOC), virtual oscillator control (VOC), grid forming converter, fault ride-through, critical clearing angle, current saturation
\end{IEEEkeywords}

\IEEEpeerreviewmaketitle

\bstctlcite{IEEEexample:BSTcontrol}

\section{Introduction}
\label{sec:introduction}
Unified virtual oscillator control (uVOC) provides a unified analysis, design, and control implementation framework for both grid-forming (GFM) and grid-following (GFL) converters \cite{uVOCorg}. The GFM operation of uVOC leverages the rigorous analytical foundation developed for dispatchable virtual oscillator control (dVOC) \cite{dVOC1,dVOC2,dVOC3,dVOC4}. In GFL applications, bidirectional power flow control and DC bus voltage regulation is achieved. In all modes of operation, no phase-locked-loop (PLL) is required which enables to circumvent the synchronization issues associated with PLLs under weak grid conditions. Moreover, fast over-current limiting and fault ride-through (FRT) is obtained without the need for switching to a back-up controller during fault. 

Implementation of over-current limiting control in GFM converters remains an open research problem till date \cite{tsaOV}. In GFM control structures employing cascaded voltage and current tracking loops, explicit current limiters on the current references offer the most intuitive solution; however, such limiters lead to saturation and eventual instability in the outer synchronization loops such as the active and reactive power control loops \cite{outerLoopSat}. Instead of imposing explicit current limits, virtual impedance for current limiting has been proposed \cite{tsaDCcl1}. As an alternative, droop controller considering current limits were proposed in \cite{tsaDCcl2,tsaDCcl5}. Explicit current limiters along with a dynamic virtual resistance was proposed in \cite{tsaDCcl4}. Contrarily, uVOC is a nonlinear time-domain controller which does not employ inner voltage and current tracking loops. Unlike droop based methods, uVOC includes current reference generation in the synchronizing oscillator, which facilitates imposing explicit current limiters without saturating the synchronization loop \cite{uVOCorg}. However, transient stability analysis is required for control design to achieve effective FRT. 

Transient stability is defined as the ability of converters to maintain synchronism with the utility grid when subjected to a major grid disturbance \cite{kundur}. For instance, a voltage source converter may experience a sudden large voltage sag due to an upstream AC fault. Extensive research on transient stability for PLL based GFL converters were reported in \cite{tsaGFL1,tsaGFL2}. Transient stability of droop and/or virtual synchronous machine (VSM) based grid forming converters were reported in \cite{tsaDC1,tsaDC2,tsaVSG,tsaPSC1}. For ease of analysis and modelling most transient stability studies ignore the converter current limit imposed by hardware capability\cite{tsaDC1,tsaDC2,tsaVSG}. In\cite{tsaDVOC}, transient stability of dVOC is studied ignoring converter current limits. Critical clearing angle (CCA) defines a characteristic limitation of second order controllers emulating virtual inertia such as droop controllers and VSM \cite{tsaDC1}; synchronization stability is lost if a fault without an equilibrium is not cleared before the power angle exceeds the CCA. Due to its first order nature, dVOC was shown to be unconstrained by such a CCA under current unsaturated faults \cite{tsaDVOC}. However, most severe fault events cause current saturation which leads to significantly distinctive dynamic response from that under current-unsaturated conditions. Moreover, power angle ($\delta$) dynamics and voltage ($V$) dynamics are typically studied separately; for instance, $\Dot{\delta}-\delta$ plots are used in most transient stability studies \cite{tsaDC1,tsaDC2,tsaVSG,tsaPSC1,tsaDVOC}. In addition, separate $V-\delta$ curves are used, where the voltage dynamics $\Dot{V}$ is implicit. Consequently, such analyses does not provide comprehensive insight into the overall system dynamics. Evidently, the state-of-the-art lacks effective analysis method for transient stability assessment of oscillator based controllers. 

The key contributions of this work are twofold: Introduction of a convenient phase-plane analysis based graphical method using $\Dot{x}-x$ surfaces which incorporates both $\Dot{V}$ and $\Dot{\delta}$ dynamics in a single three-dimensional figure; and development of a comprehensive modelling and analysis methodology for assessing transient stability of uVOC based converters under both current-constrained and unconstrained fault conditions by leveraging the $\Dot{x}-x$ surfaces. The rest of the paper is organized as follows: First, uVOC structure with FRT features is presented. Second, illustrative examples of current-saturated and unsaturated faults are presented and subsequently dynamic models are developed for both conditions. Third, the $\Dot{x}-x$ surfaces are introduced. Fourth, experimental results are presented to validate the modelling, methodology, and analysis.

\begin{figure}[htb]
	\makebox[\linewidth][c]{\includegraphics[angle = 0, clip, trim=0cm 0cm 0cm 0cm,  width=0.5\textwidth]{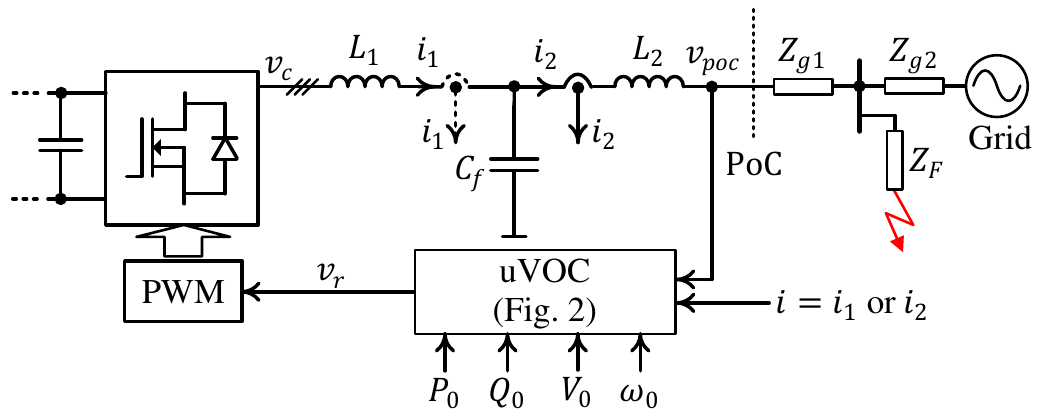}}
	\caption{A uVOC based VSC under an upstream AC fault.}
	\label{fig:sysDes}
\end{figure}

\begin{figure}[b]\setlength{\hfuzz}{1.1\columnwidth}
\begin{minipage}{\textwidth}
\makebox[\linewidth][c]{\includegraphics[angle = 0, clip, trim=0cm 0cm 0cm 0cm,  width=1\textwidth]{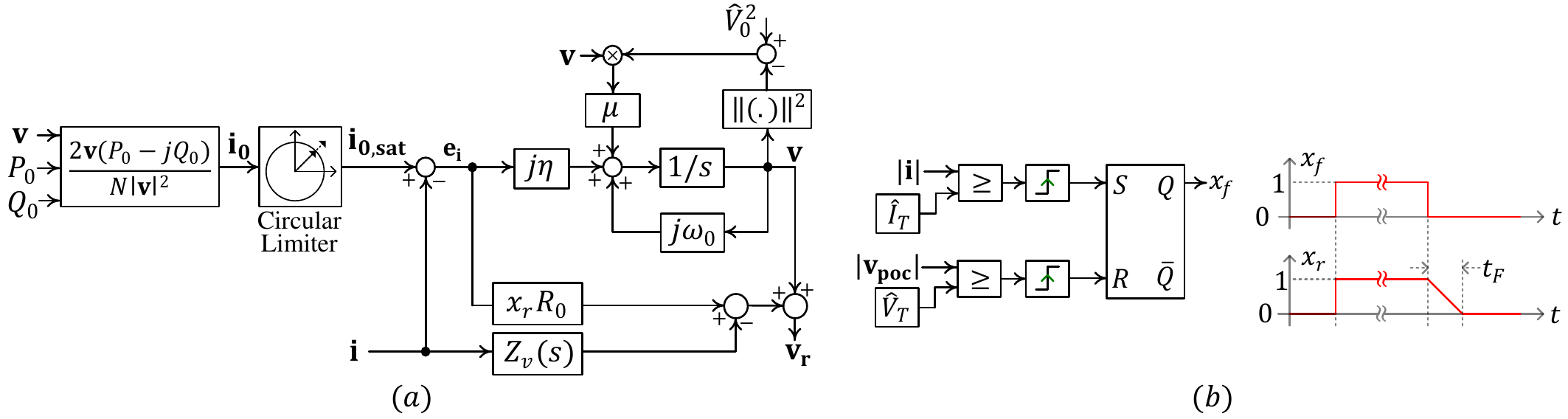}}
	\caption{uVOC structure - (a) controller implementation, (b) fault management sub-system generates two signals $x_f$ and $x_r$ which are used for transitioning between normal and current-constrained modes of operation.}
	\label{fig:uVOC}
\end{minipage}
\end{figure}
\section{Converter and Control Structure}\label{sec:sysDes}
A VSC equipped with an LCL filter employing uVOC is depicted in Fig.~\ref{fig:sysDes}, whereas the uVOC structure is shown in Fig.~\ref{fig:uVOC}. The identical uVOC controller structure is used for FRT as well. Although uVOC can be configured to operate in both grid-forming and grid-following modes under normal conditions, in this work only the grid-forming operation with real power vs. frequency and reactive power vs. voltage droop is considered for brevity. Additionally, only symmetrical faults are considered. The transient response of the system under extreme grid events such as an upstream fault is of interest for proper design of the controller to achieve effective FRT. To emulate such faults of varying severity while looking into the point of coupling (PoC), the electrical network is modelled with a fault through an impedance $Z_f$ (see Fig.~\ref{fig:sysDes}). Any such fault leads to a fast and proportionate voltage drop at the PoC leading to a large-signal response by the VSC.
Prior to delving into the transient stability assessment, the controller structure is briefly outlined in the following subsection.         

\subsection{Controller (uVOC) Structure}
uVOC is a nonlinear time-domain controller implemented in the stationary $\alpha \beta$ reference frame. In the analysis that follows, boldface notation is used to indicate space vectors in $\alpha \beta$ frame; complex vector and column vector notations such as $\mathbf{i}=i_{\alpha}+j i_{\beta}=[i_\alpha\ i_{\beta}]^T \leftrightarrow [i_a\ i_b\ i_c]^T$, are used interchangeably. As shown in Fig.~\ref{fig:uVOC}, uVOC is a first order controller and the controller state $\mathbf{v}$ is updated following the control law given as 

\begin{equation}\label{eq:uVOC}
  \begin{split}
    \Dot{\mathbf{v}} = j\omega_0 \mathbf{v} + j\eta (\mathbf{i_{0,sat}} - \mathbf{i}) + \mu (\hat{V}^2_{0} - \lVert\mathbf{v}\rVert^2)\mathbf{v},
  \end{split}
\end{equation}

\noindent
where $j=\sqrt{-1}$ denotes the imaginary unit and $\lVert\mathbf{v}\rVert = \sqrt{v^2_\alpha+v^2_\beta}$ denotes Euclidean norm; $\omega_0$ and $\hat{V}_{0}$ denote the nominal set-points for system frequency and the L-N peak voltage, respectively. Note that $\hat{V}_0=\sqrt{2}V_0$, where $V_0$ denotes the corresponding L-N root-mean-square (RMS) value and such notation is followed for other variables throughout the rest of the paper. The voltage magnitude correction gain and synchronization gains are denoted as $\mu$ and $\eta$, respectively. The converter-side current $i_1=[i_{1,a}\ i_{1,b}\ i_{1,c}]^T$ or the grid-side current $i_2=[i_{2,a}\ i_{2,b}\ i_{2,c}]^T$ can be used as the converter output current feedback $\mathbf{i}$ for uVOC implementation. Following the instantaneous power theory, the current reference $\mathbf{i_0}$ is generated as 

\begin{equation}\label{eq:etaMu}
  \begin{split}
    \mathbf{i_0} = \frac{2(P_0 - j Q_0)\mathbf{v}}{N \lVert\mathbf{v}\rVert^2},
  \end{split}
\end{equation}

\noindent
where $P_0$ and $Q_0$ denote the references/set-points for real and reactive power, respectively and $N=1$ or $N=3$ is used for single-phase and three-phase applications, respectively. 
\enlargethispage{-16.5\baselineskip} 
\subsection{Current-Constrained Operation}
The VSC enters current-constrained operation once the converter output current $\lVert\mathbf{i}\rVert$ exceeds an over-current threshold $\hat{I}_T$, which may be caused by a fault or severe voltage-sag on the grid side. The current-limiting subsystem, shown in Fig.~\ref{fig:uVOC}(b), generates and latches a control signal $x_f$ once such an over-current event is detected. Once the terminal voltage magnitude $\lVert \mathbf{v_{poc}}\rVert$
returns above the under-voltage threshold $\hat{V}_T$, $x_f$ is cleared and the controller resumes normal operation. Another control signal $x_r$ is generated following $x_f$; $x_r$ is ramped down over the duration of $t_F$ to facilitate smooth transition from current-constrained to normal operation.

For effective FRT under current-constrained condition, two key objectives must be achieved - 

\subsubsection{Fast Over-Current Limiting}
The converter output current must be limited without saturating/destabilizing the synchronizing loop. The oscillator dynamics given by \eqref{eq:uVOC} achieves power synchronization though an instantaneous droop response \cite{uVOCorg}, which must be unaffected by the current limiting operation. Fortunately, uVOC uses explicit current references in its synchronization structure. Therefore, a circular saturation/limiter can be directly incorporated in the synchronizing controller as 

\begin{equation}\label{eq:cirLim}
  \begin{split}
    \mathbf{i_{0,sat}} = 
    \begin{cases}
        \mathbf{i_0}, & \lVert\mathbf{i_0}\rVert\leq \hat{I}_{m}\\
        \mathbf{i_0}\times (\hat{I}_m/\lvert\mathbf{i_0}\rVert), & \text{otherwise}
    \end{cases},
  \end{split}
\end{equation}

\noindent
where $I_m$ denotes the maximum allowable current constrained by the converter hardware. In Section~\ref{sec:dynModSatCase}, the circular limiter is shown to enable retention of the instantaneous droop response under current constrained operation. However, the oscillator dynamics is not typically fast enough to enable fast current limiting; an active resistance $R_0$ is used on the current error $(\mathbf{i_{0,sat}}-\mathbf{i})$ for faster current limiting response. As can be seen in Fig.~\ref{fig:uVOC}, the synchronization gain $\eta$ serves as a complex integral gain on the current error and the integral action can be augmented by raising the gain as $\eta =\eta_0 (1+x_r/\tau_f)$, where smaller $\tau_f$ leads to faster settling time.


\subsubsection{Feasibility Constraint}
\begin{figure}[htb]
	\makebox[\linewidth][c]{\includegraphics[angle = 0, clip, trim=0cm 0cm 0cm 0cm,  width=0.325\textwidth]{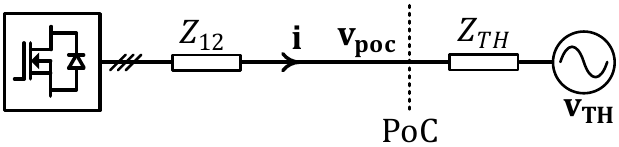}}
	\caption{Equivalent system model for analysis.}
	\label{fig:equiCkt}
\end{figure}

The voltage magnitude correction term in \eqref{eq:uVOC}, tends to restore the nominal voltage magnitude which may not be feasible in case of severe grid faults/voltage sags under current constrained operation. The feasibility constrain can be explained using a simplified system model. For clarity and ease of analysis, the system shown in Fig.~\ref{fig:sysDes} can be simplified into an equivalent representation shown in Fig.~\ref{fig:equiCkt}. In the frequency range of interest for transient stability assessment, the control-computation delay, effect of the PWM, and the filter capacitor $C_f$ can be ignored~\cite{uVOCorg,psc_org,tsaDC1}. The converter-side and grid-side inductors and the corresponding parasitic resistances are combined into $Z_{12} = s(L_1+L_2)+R_1+R_2$. The grid/rest of the network can be modelled by a Thevenin equivalent. The grid impedance $Z_{g1}$ and $Z_{g2}$, location of the fault, and the impedance $Z_f$ (see Fig.~\ref{fig:sysDes}) result in the equivalent source $\mathbf{v_{TH}}$ and equivalent impedance $Z_{TH}$, seen by the VSC while looking into the PoC. By varying the equivalent source and impedance, faults of varying severity can be emulated in the analysis and during experiments.     

\noindent
To illustrate the feasibility constraint, we consider a grid fault with $v_{TH}=0.5$ p.u. and $Z_{TH}= 0.1$ p.u.; the converter terminal voltage cannot be regulated to larger than $0.62$ p.u. if the output current is constrained at $1.2$ p.u. during the fault. Such physical limits must be respected during FRT. Therefore, the voltage magnitude correction gain is parameterized as $\mu = (1-x_r) \mu_0$, which disables the magnitude correction term under current constrained operation. Unlike droop control or VSM, no inner voltage or current reference tracking loops are used in uVOC; the uVOC output $\mathbf{v_r}$, given as 

\begin{equation}\label{eq:vr}
  \begin{split}
    \mathbf{v_r} = \mathbf{v} + R_0(\mathbf{i_{0,sat}} - \mathbf{i}) - Z_v(s) \mathbf{i},
  \end{split}
\end{equation}

\noindent
is used directly by the pulse-width-modulator (PWM). The virtual impedance $Z_v(s)$ is used for stabilization of the oscillator\cite{uVOCorg} and harmonic voltage/current compensation\cite{TpelHCS,hvsApec}.

\section{Faults Under Current-Constrained and Unconstrained Conditions}
Potential fault events can be classified into two categories - faults under unsaturated current conditions and those under saturated current conditions. To illustrate how a fault may occur without exceeding the converter current limit, first, we briefly review the droop characteristic of uVOC given as 

\begin{equation}\label{eq:dr}
  \begin{split}
    \Dot{V} &= 2\mu V (V^2_0 - V^2) + \frac{\eta}{N V}(Q_0 - Q),\\
    \omega  &= \omega_0 + \frac{\eta}{N V^2}(P_0 - P),
  \end{split}
\end{equation}

\noindent
where $\mathbf{v}=\sqrt{2}Ve^{j \omega t}$ and $P$ and $Q$ denote the real and reactive power output, respectively. The detail derivation can be found in \cite{uVOCorg}. Evidently, the controller exhibits $Q-V$ and $P-\omega$ droop response. In an event of a voltage sag caused by an up-stream fault, the controller increases its reactive power output to reduce the voltage deviation from the nominal set-point $V_0$. Based on the severity of the fault and other operating conditions such as grid frequency and power set-points $P_0$ and $Q_0$, the converter may reach steady-state without reaching the over-current limit. In such cases, the converter does not enter the current-constrained operation marked by $x_f=x_r=1$. However, in case of more severe faults, the converter may enter current-constrained operation. To illustrate the effect of such fault events, we consider a system described in Appendix~\ref{apndx:simExpSetup}. Examples of two types of faults are described in the following subsections.

\begin{figure}[htb]
	\makebox[\linewidth][c]{\includegraphics[angle = 0, clip, trim=0cm 0cm 0cm 0cm,  width=0.45\textwidth]{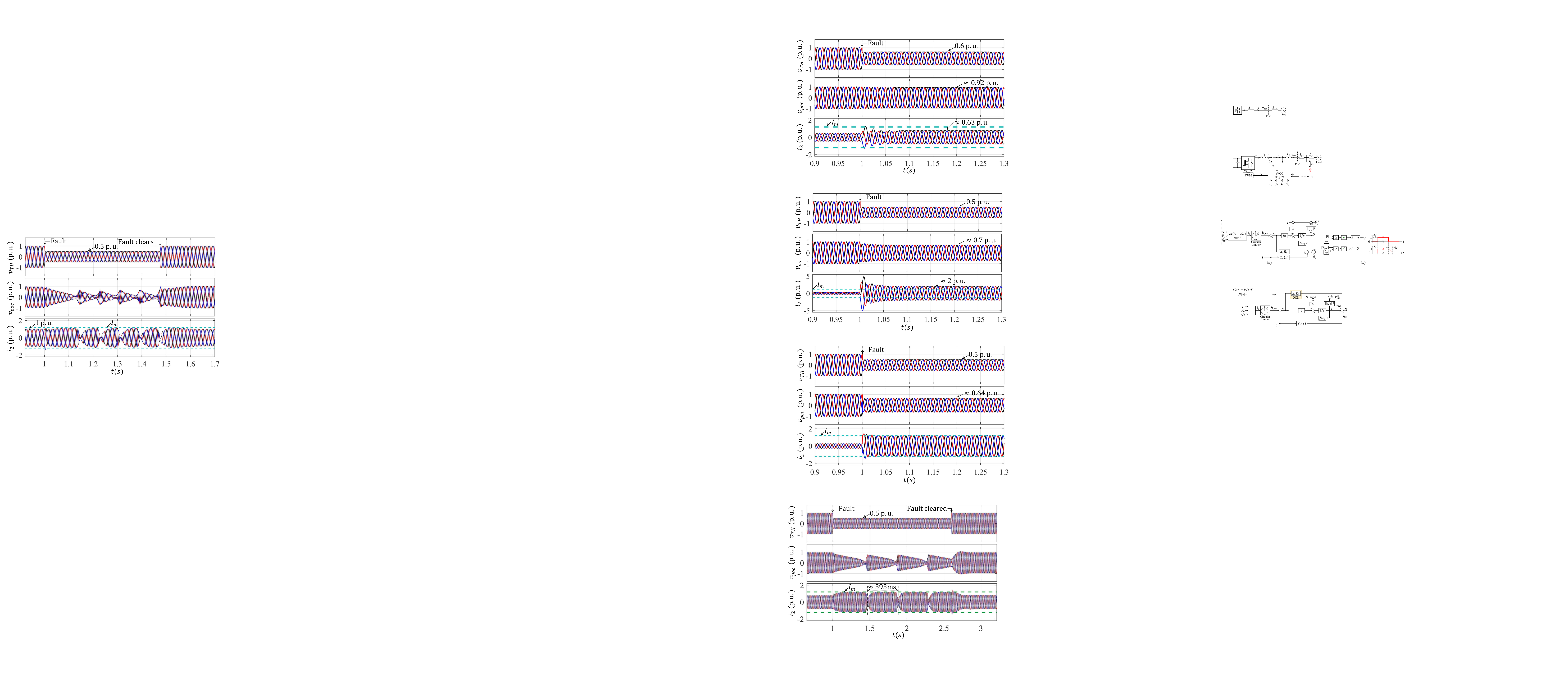}}
	\caption{Simulated fault with $v_{TH} = 0.6\ \text{p.u.}$ and $Z_{TH} \approx 0.52\ \text{p.u.}$ when converter output current does not reach the maximum limit.}
	\label{fig:simUnsatFLT}
\end{figure}

\begin{figure}[htb]
	\makebox[\linewidth][c]{\includegraphics[angle = 0, clip, trim=0cm 0cm 0cm 0cm,  width=0.45\textwidth]{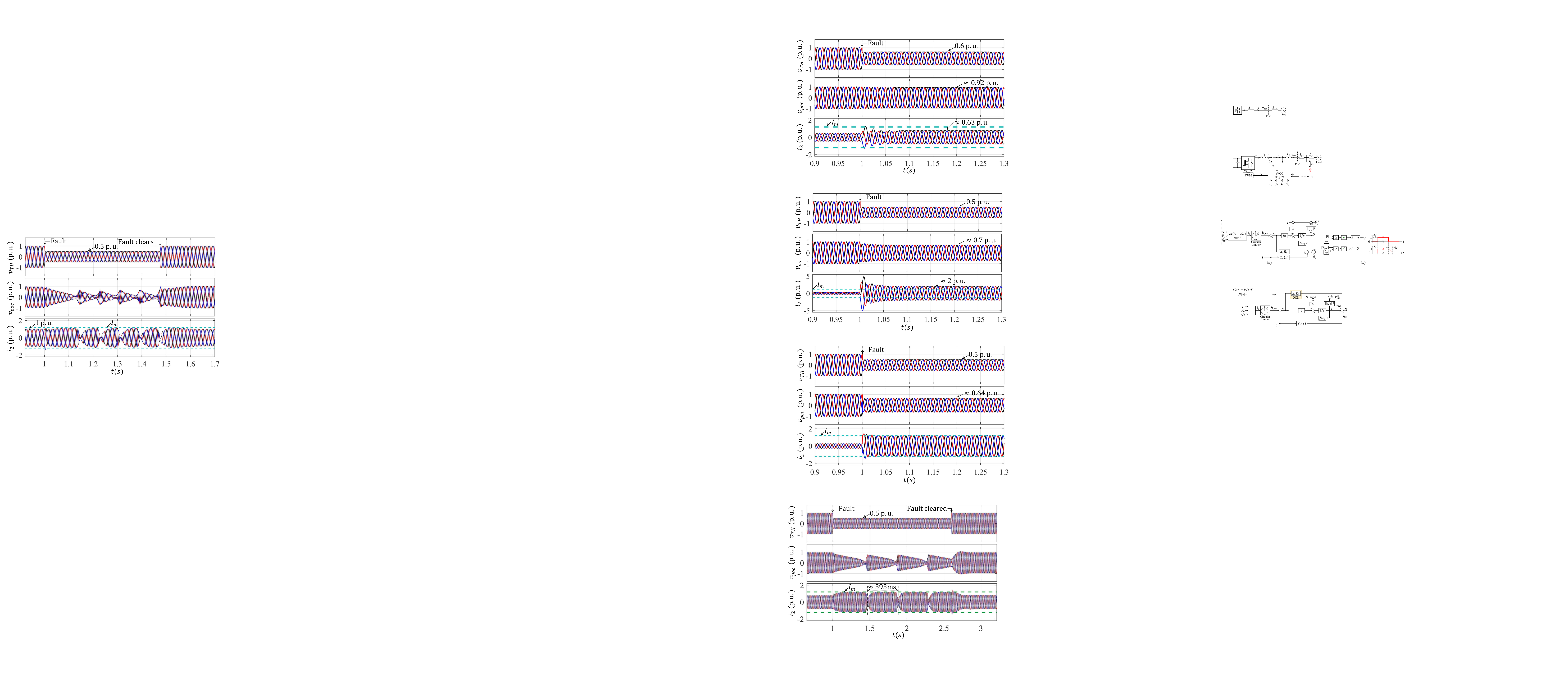}}
	\caption{Simulated fault with $v_{TH} = 0.5\ \text{p.u.}$ and $Z_{TH} \approx 0.1\ \text{p.u.}$ with current-limiting sub-system and circular limiter disabled.}
	\label{fig:simSatFLT}
\end{figure}

\subsection{Case I : Current-Unconstrained Fault}\label{sec:simCaseI} 
A simulated grid fault under current-unconstrained condition for a three-phase VSC is shown in Fig.~\ref{fig:simUnsatFLT}; the simulation is performed using detailed switching model in \emph{PLECS Standalone} platform. The equivalent grid impedance is taken as $Z_{TH} \approx 0.52\ \text{p.u.}$ with an reactance-to-resistance ($X/R$) ratio of $20$ which gives a short-circuit ratio (SCR) of $\approx 1.9$. The power references are set as $P_0 = 0.38\ \text{p.u.}$ and $Q_0 = 0$; a fault is emulated by introducing a step voltage sag from $v_{TH} = 1\ \text{p.u.}$ to $v_{TH} = 0.6\ \text{p.u.}$ (see Fig.~\ref{fig:equiCkt}). Due to its grid-forming nature, the converter increases its reactive power output which is reflected in the output current, i.e., $i_2$ increases to $0.63\ \text{p.u.}$; consequently, the PoC voltage is raised to $0.92\ \text{p.u.}$. The converter output current $i_2$ does not exceed the current limit $I_m = 1.2\ \text{p.u.}$ during the initial overshoot as well as when steady-state is reached.

\subsection{Case II : Current-Constrained Fault}\label{sec:simCaseII}
To model a more severe fault, grid impedance of $Z_{TH} = 0.1\ \text{p.u.}$ is considered. The current-limiting subsystem, shown in Fig.~\ref{fig:uVOC}(b), is disabled and the circular limiter, shown in shown in Fig.~\ref{fig:uVOC}(a), is excluded to illustrate how the converter would respond without any current-limit constrains. The power references are set as $P_0 = 0.27\ \text{p.u.}$ and $Q_0 = 0$. Now the fault is emulated by a step change from $v_{TH} = 1\ \text{p.u.}$ to $v_{TH} = 0.5\ \text{p.u.}$; although the PoC voltage is raised to $0.7\ \text{p.u.}$, the output current goes through a large transient exceeding the maximum limit $I_m$ and settles at $i_2\approx 2\ \text{p.u.}>I_m$. Evidently, the current-limit is exceeded by the initial overshoot. Without an effective current-limiting and FRT controller, the VSC must disconnect from the grid/electrical network under such fault conditions to protect the hardware. Therefore, an appropriate FRT controller needs to be implemented and comprehensive transient stability analysis is required for the design of the FRT controller. The current-limiting subsystem and the circular limiter block are responsible for clamping the output current at $I_m$, once such an over-current is detected.

The converter and the control system exhibit distinctly different dynamic responses for the two different cases. Effective FRT requires proper analysis and design for transient stability of the system in both cases. In the following section, dynamic models are developed for such analyses.

\section{Dynamic Model}
Without loss of generality, the dynamic model is developed in a synchronous reference frame rotating at the grid frequency $\omega_g$ and aligned with the equivalent grid voltage $\mathbf{v_{TH}}=\sqrt{2}V_{TH}e^{j\omega_g}$. 

\subsection{Current-Unconstrained Operation}
During current-unconstrained operation, the active resistance is disabled and $\mu\neq 0$, and $\mathbf{i_{0,sat}}=\mathbf{i_0}$. The voltage magnitude and angle dynamics, given by \eqref{eq:dr}, can be obtained in the synchronous reference frame as

\begin{equation}\label{eq:drSynRF}
  \begin{split}
    \Dot{V} &= 2\mu V (V^2_0 - V^2) + \frac{\eta}{N V}(Q_0 - Q),\\
    \Dot{\delta}  &= \omega_0 - \omega_g + \frac{\eta}{N V^2}(P_0 - P),
  \end{split}
\end{equation}

\noindent
where $\int \omega dt = \omega_g t + \delta$. Note that $\delta$ is termed as the so-called \emph{power angle}. The dynamics of the converter output current $\mathbf{i_{dq}}=i_d + j i_q \leftrightarrow \sqrt{2}(I_d+j I_q)$ in the synchronous frame can be derived as

{\small
\begin{equation}\label{eq:idqUnsatFLT}
  \begin{split}
    \left[{\begin{array}{c}
         \Dot{I}_d\\
         \Dot{I}_q\\
         \end{array}}\right]=\left[{\begin{array}{cc}
         -\frac{R_{e}}{L_e} & \omega_g\\
         -\omega_g & -\frac{R_{e}}{L_e}\\
         \end{array}}\right]\left[{\begin{array}{c}
         I_d\\
         I_q\\
         \end{array}}\right]+\frac{1}{L_e}\left[{\begin{array}{c}
         V\cos{(\delta)} - V_g\\
         V\sin{(\delta)}\\
         \end{array}}\right].
  \end{split}
\end{equation}
}

\noindent
Here, the equivalent resistance $R_e=R_v+R_1+R_2+R_{TH}$ and the equivalent inductance $L_e=L_v+L_1+L_2+L_{TH}$, whereas the virtual impedance $Z_v(s)$ (see Fig.~\ref{fig:uVOC}) is implemented as 

\begin{equation}\label{eq:Zv}
  \begin{split}
    Z_v(s) = \frac{R_{v0}}{s/\omega_b+1}+x_r\frac{s L_{v0}}{s/\omega_b+1}.
  \end{split}
\end{equation}

\noindent
The bandwidth $\omega_b$ is selected so as to avoid amplification of high-frequency noise. Due to the limited bandwidth $\omega_b$, the effective virtual impedances differ from $L_{v0}$ and $R_{v0}$ and can be determined as $L_v\approx\operatorname{Im}\{Z_v(j\omega_0)\}$ and $R_v\approx\operatorname{Re}\{Z_v(j\omega_0)\}$. Note that the virtual inductor is not typically required during normal operation but rather used for limiting the current overshoot during the initial fault instants and fault-clearing. The first-order filter corresponding to the virtual inductance in \eqref{eq:Zv} is updated by the digital controller irrespective of mode of operation but the filter output is used with the scaling of $x_r$. Setting $\Dot{I}_d=0, \Dot{I}_q=0$ and taking $X_e=\omega_g L_e$, the output current components are obtained and the real and reactive power outputs are derived as 


\begin{equation}\label{eq:pqUnsat}
  \begin{split}
    P &= \frac{N[V^2 R_e - V V_{TH} \{R_e \cos{(\delta)}-X_e \sin{(\delta)}\}]}{R^2_e+X^2_e},\\
    Q &= \frac{N[V^2 X_e - V V_{TH} \{X_e \cos{(\delta)}+R_e \sin{(\delta)}\}]}{R^2_e+X^2_e}.
  \end{split}
\end{equation}

\begin{figure}[htb]
	\makebox[\linewidth][c]{\includegraphics[angle = 0, clip, trim=0cm 0cm 0cm 0cm,  width=0.425\textwidth]{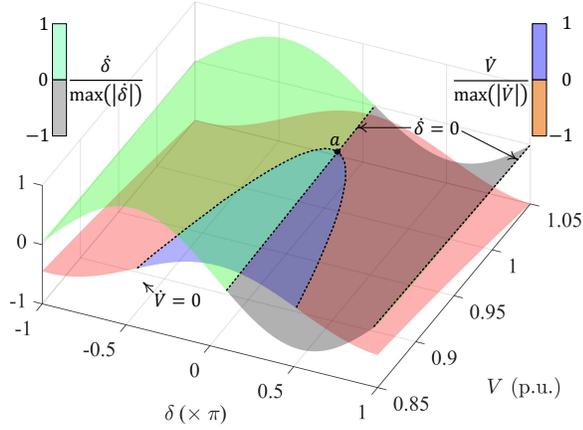}}
	\caption{Normalized $\Dot{x}-x$ surfaces at pre-fault condition for \emph{Case I}.}
	\label{fig:xDotxPreFLT_CaseI}
\end{figure}

\begin{figure}[b]\setlength{\hfuzz}{1.1\columnwidth}
\begin{minipage}{\textwidth}
\makebox[\linewidth][c]{\includegraphics[angle = 0, clip, trim=0cm 0cm 0cm 0cm,  width=0.85\textwidth]{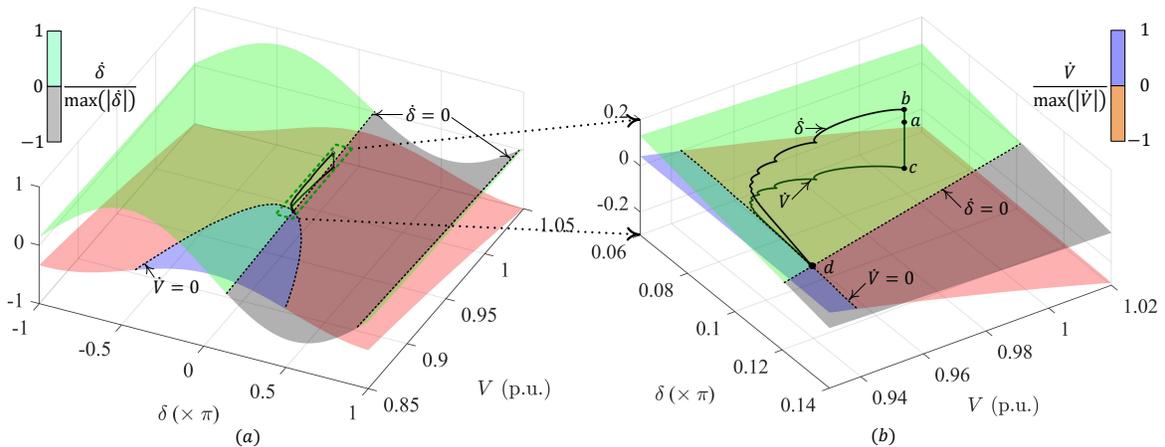}}
	\caption{Normalized $\Dot{x}-x$ surfaces under current unsaturated fault condition (\emph{Case I}) - (a) overall $\Dot{x}-x$ surfaces, (b) zoomed-in view showing the movement of operating point from pre-fault equilibrium $a$ to equilibrium $d$ under fault condition.}
	\label{fig:xDotxFLT_CaseI}
\end{minipage}
\end{figure}

\subsection{Current-Constrained Operation}\label{sec:dynModSatCase}
Under current-constrained operation, the active resistance is enabled and $\mu=0$ and the saturated current reference $\mathbf{i_{0,sat}}$ is given as

\begin{equation}\label{eq:I0satab}
  \begin{split}
    \mathbf{i_{0,sat}}=\frac{\mathbf{i_0}}{\lVert \mathbf{i_0}\rVert}\times \hat{I}_m \Rightarrow \mathbf{i_{0,sat}} = \frac{\hat{I}_m(P_0 - j Q_0)\mathbf{v}}{S_0 \lVert \mathbf{v} \rVert},
  \end{split}
\end{equation}

\noindent
where $S_0=\sqrt{P^2_0+Q^2_0}$. Substituting \eqref{eq:I0satab} into \eqref{eq:uVOC}, the voltage magnitude and angle dynamics can be derived as 

\begin{equation}\label{eq:drSynRFsatFLT}
  \begin{split}
    \Dot{V} &= \frac{\eta}{N V}\left(\frac{N V I_m}{S_0}Q_0 - Q\right),\\
    \Dot{\delta}  &= \omega_0 - \omega_g + \frac{\eta}{N V^2}\left(\frac{N V I_m}{S_0}P_0 - P\right).
  \end{split}
\end{equation}

\noindent
Evidently, the power synchronization mechanism is retained under current-constrained operation since the $\Dot{\delta}-P$ dynamics is unaltered; only the power reference set-points are scaled proportionately with the voltage magnitude. In synchronous reference frame, the saturated current reference $\mathbf{i_{0,sat_{dq}}}=\sqrt{2}(I_{d0}+j I_{q0})$ is given as 

\begin{equation}\label{eq:I0sat}
  \begin{split}
    \left[{\begin{array}{c}
         I_{d0}\\
         I_{q0}\\
         \end{array}}\right]=\frac{I_m}{S_0}\left[{\begin{array}{c}
         P_0 \cos{(\delta)}+Q_0 \sin{(\delta)}\\
         P_0 \sin{(\delta)}-Q_0 \cos{(\delta)}\\
         \end{array}}\right]
  \end{split}
\end{equation}

\noindent
The converter output current dynamics can be derived as 



{\small
\begin{equation}\label{eq:idqSatFLT}
  \begin{split}
    \left[{\begin{array}{c}
         \Dot{I}_d\\
         \Dot{I}_q\\
         \end{array}}\right]=\left[{\begin{array}{cc}
         -\frac{R_{e}}{L_e} & \omega_g\\
         -\omega_g & -\frac{R_{e}}{L_e}\\
         \end{array}}\right]&\left[{\begin{array}{c}
         I_d\\
         I_q\\
         \end{array}}\right]\\+\frac{1}{L_e}&\left[{\begin{array}{c}
         V\cos{(\delta)} + R_0 I_{d0} - V_g\\
         V\sin{(\delta)} + R_0 I_{q0}\\
         \end{array}}\right].
  \end{split}
\end{equation}
}
\noindent
The real and reactive power outputs are obtained as 

{\small
\begin{equation}\label{eq:pSat}
  \begin{split}
    P = \frac{N}{R^2_e+X^2_e}[V^2 R_e - V V_{TH} \{R_e \cos{(\delta)}-X_e \sin{(\delta)}\}\\+(V R_0 I_m/S_0)(R_e P_0 - X_e Q_0)],\\
  \end{split}
\end{equation}
\begin{equation}\label{eq:qSat}
  \begin{split}
    Q = \frac{N}{R^2_e+X^2_e}[V^2 X_e - V V_{TH} \{X_e \cos{(\delta)}+R_e \sin{(\delta)}\}\\+(V R_0 I_m/S_0)(X_e P_0 + R_e Q_0)].
  \end{split}
\end{equation}
}

\section{Transient Stability Assessment}
In \cite{tsaDC1}, for transient stability analysis of droop based converters $\Dot{\delta}-\delta$ and $V-\delta$ curves are used where the voltage dynamics $\Dot{V}$ is implicit and the system equilibrium points over $V-\delta$ distribution cannot be determined through the graphical analysis. uVOC also exhibits dynamic behavior in both the power angle $\delta$ and the voltage magnitude $V$, as shown by \eqref{eq:drSynRF} and \eqref{eq:drSynRFsatFLT}. Consequently, $\Dot{\delta}-\delta$ and/or $V-\delta$ curves do not show a holistic response of the system and hence are not sufficient for transient stability assessment of uVOC. Therefore, for a comprehensive analysis we adopt superimposed three-dimensional plots of $\Dot{\delta}$ and $\Dot{V}$ over the distribution of $V$ and $\delta$, denoted as $\Dot{x}-x$ surfaces henceforth.  

\begin{figure*}[tb]
	\makebox[\linewidth][c]{\includegraphics[angle = 0, clip, trim=0cm 0cm 0cm 0cm,  width=0.85\textwidth]{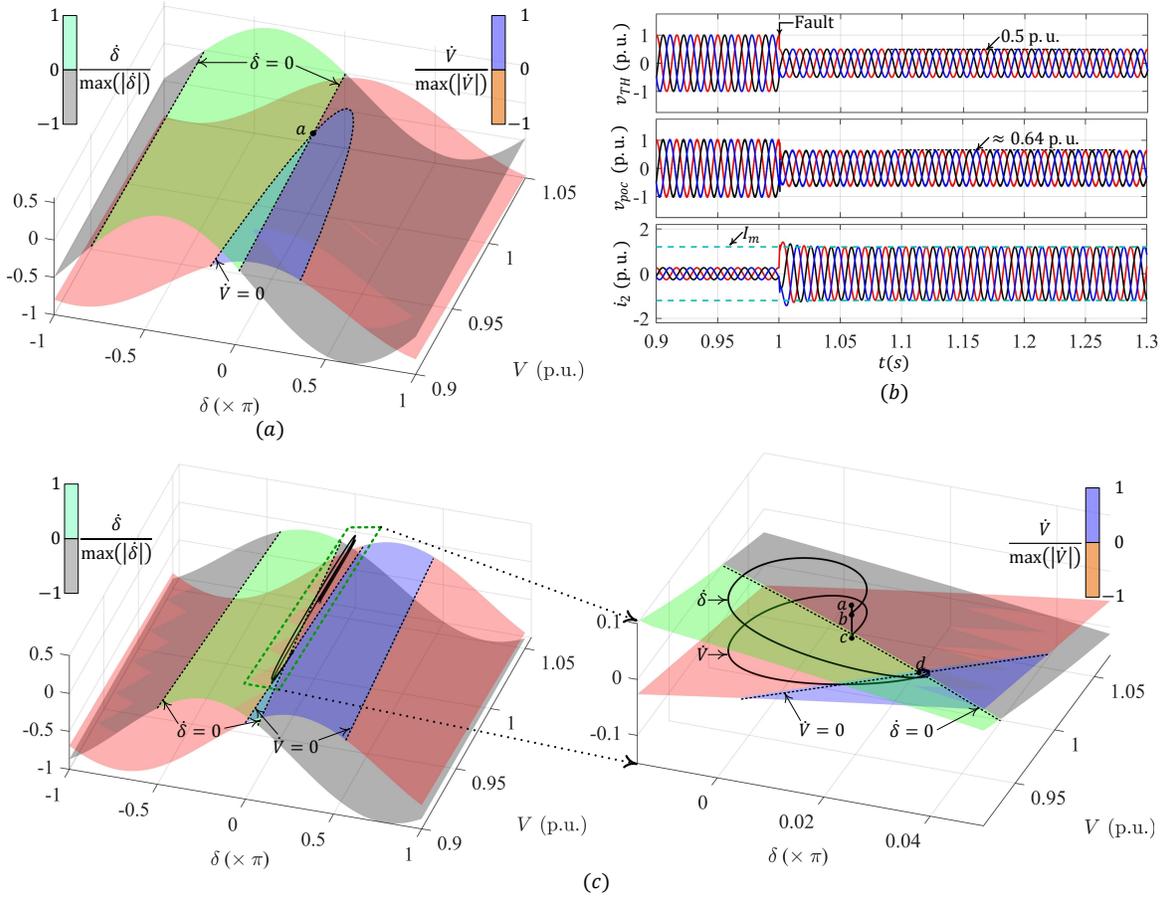}}
	\caption{Phase-plane analysis and simulated response for current-constrained fault (\emph{Case II}) - (a) normalized $\Dot{x}-x$ surfaces at pre-fault condition, (b) simulated current-constrained fault with $v_{TH} = 0.5\ \text{p.u.}$ and $Z_{TH} \approx 0.1\ \text{p.u.}$ with current-limiting sub-system and circular limiter enabled, (c) normalized $\Dot{x}-x$ surfaces and movement of operating point from pre-fault equilibrium $a$ to equilibrium $d$ under fault condition.}
	\label{fig:xDotxFLT_CaseII}
\end{figure*}

\begin{figure*}[tb]
	\makebox[\linewidth][c]{\includegraphics[angle = 0, clip, trim=0cm 0cm 0cm 0cm,  width=0.85\textwidth]{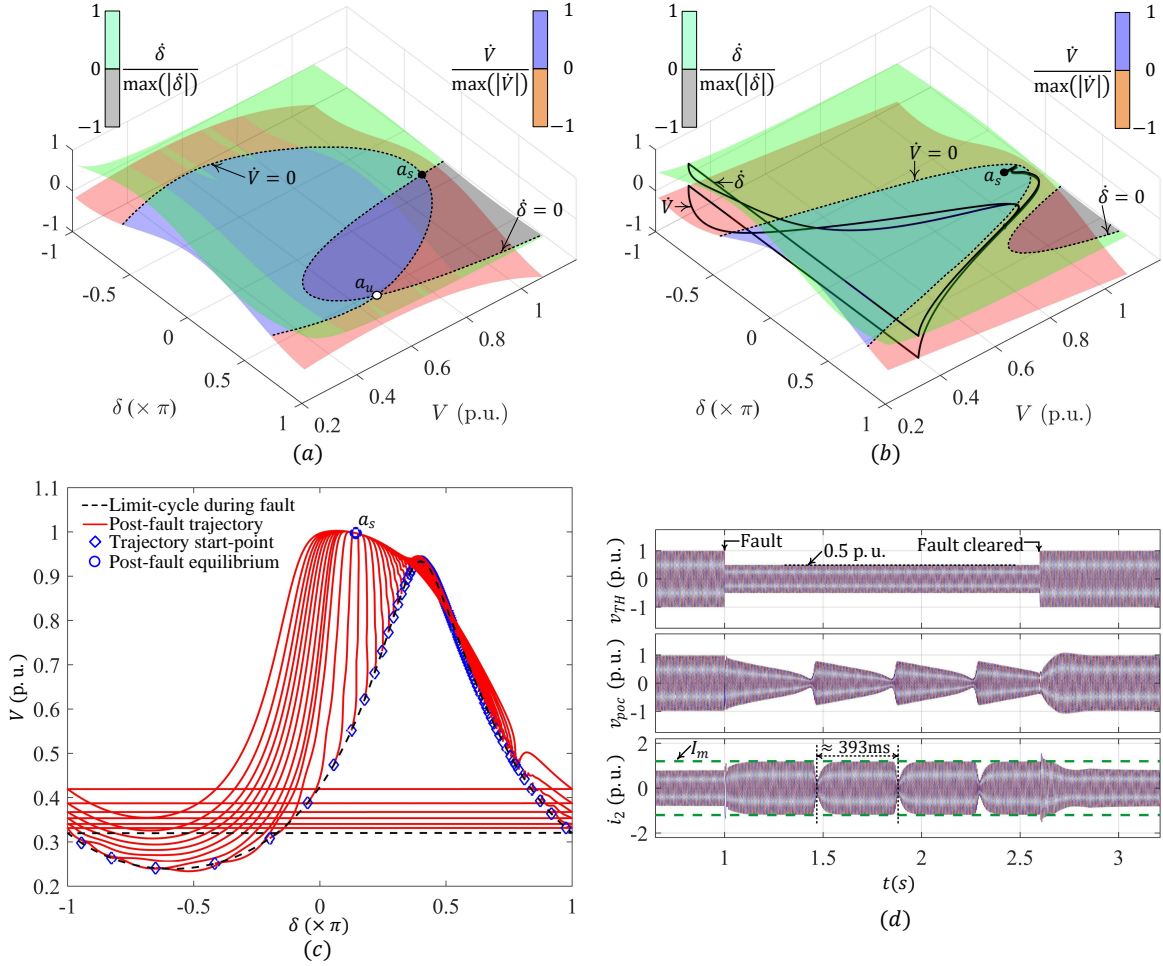}}
	\caption{Phase-plane analysis and simulated response for fault without an equilibrium (\emph{Case III}) - (a) normalized $\Dot{x}-x$ surfaces at pre-fault condition, (b) normalized $\Dot{x}-x$ surfaces at fault condition, when no equilibrium exists and the system trajectory follows a limit-cycle, (c) system trajectory after fault clears at different points on the limit-cycle; all trajectories reach the post-fault equilibrium, (d) simulated fault response when no equilibrium exists during fault.}
	\label{fig:xDotxFLT_CaseIII}
\end{figure*}

\subsection{Case I : Current-Unconstrained Fault}\label{sec:tsaCaseI}
First, we consider the current-unconstrained fault described in Section~\ref{sec:simCaseI}. The $\Dot{x}-x$ surfaces at pre-fault condition are shown in Fig.~\ref{fig:xDotxPreFLT_CaseI}. 
\enlargethispage{-18.5\baselineskip}
The $\Dot{x}-x$ surfaces are generated evaluating the time-derivatives $\Dot{\delta}$ and $\Dot{V}$ using $\eqref{eq:drSynRF}$ over the two-dimensional distribution of $\delta$ and $V$. For a clear graphical representation, the derivative values are normalized by their respective absolute maximum values over the whole $\delta - V$ distribution. Evidently, a unique equilibrium point, marked as $a$, is found at the intersection of $\Dot{\delta}=0$ and $\Dot{V}=0$ curves. The small signal-stability of the equilibrium point $a\equiv (\delta_a,V_a)$ can be readily evaluated graphically through inserting small perturbation; for instance, while the operating point is perturbed slightly to some power angle $\delta<\delta_a$, the derivative of power angle becomes positive, i.e., $\Dot{\delta}>0$ and consequently the operating point moves back towards $a$.
Conversely, small increase in the power angle as $\delta>\delta_a$, makes the power angle derivative $\Dot{\delta}<0$ which eventually forces the operating point to move back towards $a$. Similarly, the small signal stability along the $V$-axis can be verified. 

Next, when a sudden voltage sag is introduced in the equivalent grid source from $v_{TH} = 1\ \text{p.u.}$ to $v_{TH} = 0.6\ \text{p.u.}$, as described in Section~\ref{sec:simCaseI}, the resulting $\Dot{x}-x$ surfaces are depicted in Fig.~\ref{fig:xDotxFLT_CaseI}(a). The corresponding zoomed-in view of the system trajectory, i.e., the movement of the system operating point, is shown in Fig.~\ref{fig:xDotxFLT_CaseI}(b). The system trajectory is obtained through time-domain numerical solution of the system model given by \eqref{eq:drSynRF} and \eqref{eq:idqSatFLT}; the \emph{ode45} solver \cite{dormand_prince_1980}, based on the Dormand-Prince (4,5) pair \cite{matlabODE}, was used in MATLAB to obtain the system trajectory. 

At the instant of the fault, the power angle $\delta$ and the voltage $V$ remain at the pre-fault equilibrium $(\delta_a,V_a)$, but the derivatives move to their respective $\Dot{\delta}$ and $\Dot{V}$ surfaces arriving at $b$ and $c$, respectively. The derivatives move along their respective surfaces and eventually reach a new equilibrium $d$. The movement of the system trajectory and the equilibrium points at pre-fault and fault conditions match identically with simulated results shown in Fig.~\ref{fig:simUnsatFLT}. Similar analysis can be performed for the transient response during fault-clearing which has been excluded in the interest of space. 

\subsection{Case II : Current-Constrained Fault}\label{sec:tsaCaseII}
The $\Dot{x}-x$ surfaces at the pre-fault condition for \emph{Case II} described in Section~\ref{sec:simCaseII} are depicted in Fig.~\ref{fig:xDotxFLT_CaseII}(a); a stable and unique equilibrium $a\equiv (\delta_a,V_a)$ is observed. The corresponding simulated response is shown in Fig.~\ref{fig:xDotxFLT_CaseII}(b). 

Next, a step change from $v_{TH} = 1\ \text{p.u.}$ to $v_{TH} = 0.5\ \text{p.u.}$ causes a fault condition which leads to over-current beyond $I_m$. However, unlike Section~\ref{sec:simCaseII}, the full controller shown in Fig.~\ref{fig:uVOC}(a) and (b) including the circular current limiter and the fault-management subsystem is employed. Furthermore, once a fault is detected, the reactive power set-point is increased as $Q_0=\sqrt{S^2_0-P^2_0}$ to leverage the remaining current capability of the converter to raise the terminal voltage. The corresponding $\Dot{x}-x$ surfaces and the zoomed-in view of the system trajectory are shown in Fig.~\ref{fig:xDotxFLT_CaseII}(c). Immediately at the fault instant, the system operating point stays at the pre-fault equilibrium $a\equiv (\delta_a,V_a)$. Subsequently, the derivatives $\Dot{\delta}$ and $\Dot{V}$ reaches the new equilibrium $d$ following $a-b-d$ and $a-c-d$ trajectories on their respective surfaces. Evidently, the system reaches steady state quickly (see Fig.~\ref{fig:xDotxFLT_CaseII}(b)) as predicted by the trajectory analysis and the output current is clamped at the maximum allowable value $I_m$.

Another class of fault response is resulted if no equilibrium can be reached under fault condition, i.e., no feasible power flow for given real and reactive power set-points and grid condition. Such a case is described in the following subsection.


\subsection{Case III : Fault With No-Equilibrium Point}\label{sec:tsaCaseIII}
If no feasible equilibrium exists during fault, the system exhibits a limit-cycle behavior. To illustrate such a response, during pre-fault condition we consider $v_{TH}=1$ p.u., $Z_{TH}=0.52$ p.u.; power references are set as $P_0=0.8$ p.u. and $Q_0=0$. The $\Dot{x}-x$ surfaces corresponding to the pre-fault condition are depicted in Fig.~\ref{fig:xDotxFLT_CaseIII}(a); two equilibrium points, i.e., intersection points of $\Dot{\delta}=0$ and $\Dot{V}=0$ curves, are observed which include a stable equilibrium $a_s$ and an unstable equilibrium $a_u$. The stability property of each point can be verified graphically following the same procedure described in Section~\ref{sec:tsaCaseI}. Therefore, the system operates at $a_s$ during pre-fault condition. At fault condition triggered by a step change in the grid voltage from a step change from $v_{TH} = 1\ \text{p.u.}$ to $v_{TH} = 0.5\ \text{p.u.}$ leads to the $\Dot{x}-x$ surfaces shown in Fig.~\ref{fig:xDotxFLT_CaseIII}(b). Evidently, no equilibrium point exists, i.e., no intersection between $\Dot{\delta}=0$ and $\Dot{V}=0$ curves. Consequently, starting from the pre-fault equilibrium $a_s$, the system moves into a limit-cycle behavior. From the trajectory analysis, the period of the limit-cycle is determined as $\approx 393$ ms.

\begin{figure*}[tb]
	\makebox[\linewidth][c]{\includegraphics[angle = 0, clip, trim=0cm 0cm 0cm 0cm,  width=0.95\textwidth]{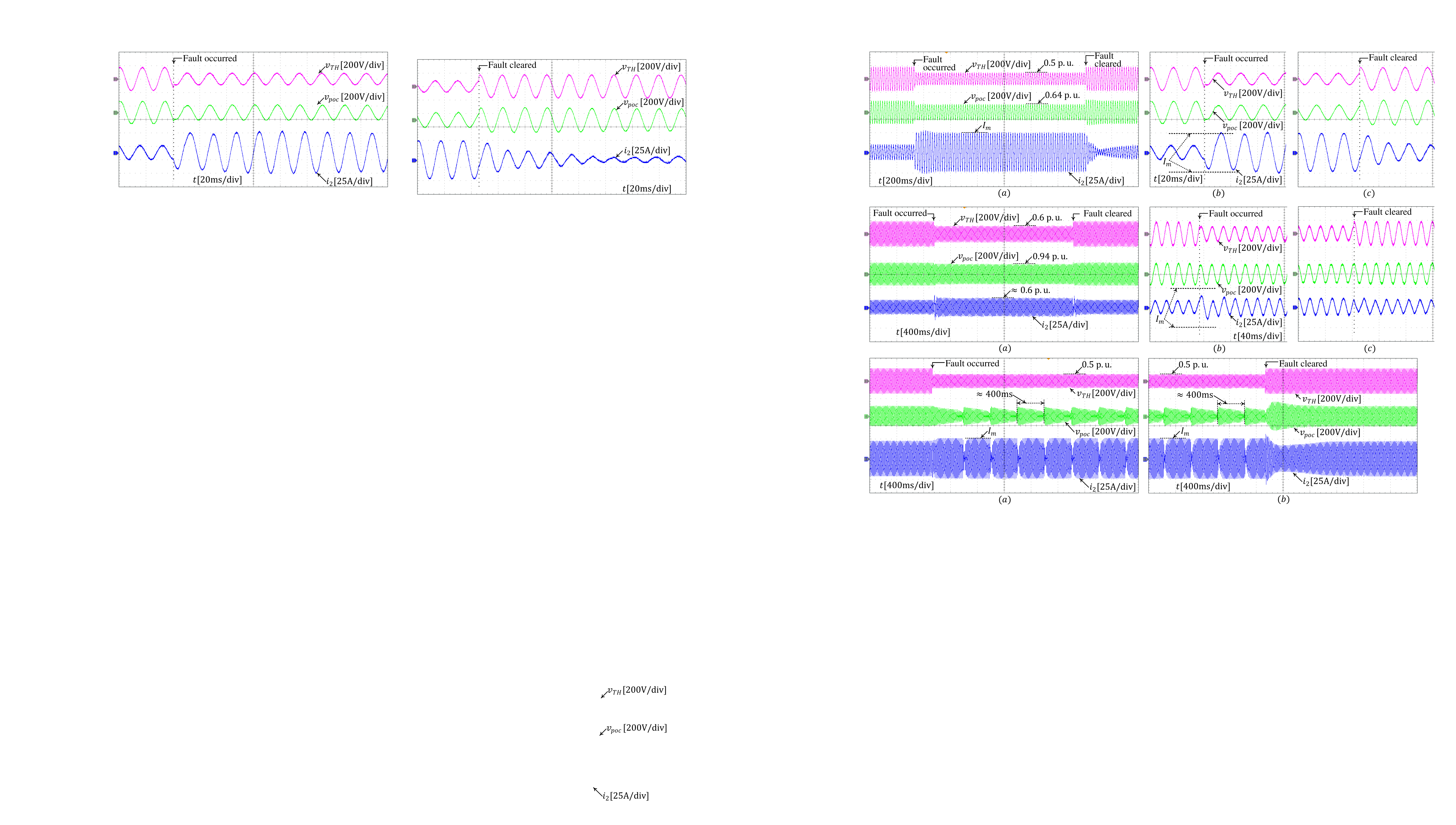}}
	\caption{Experimental results for \emph{Case I} - (a) converter response to an upstream grid fault when converter current does not reach the hardware limit, (b) zoomed-in response when fault occurs, (c) zoomed-in response during fault clearing.}
	\label{fig:exp_CaseI}
\end{figure*}

\begin{figure*}[tb]
	\makebox[\linewidth][c]{\includegraphics[angle = 0, clip, trim=0cm 0cm 0cm 0cm,  width=0.95\textwidth]{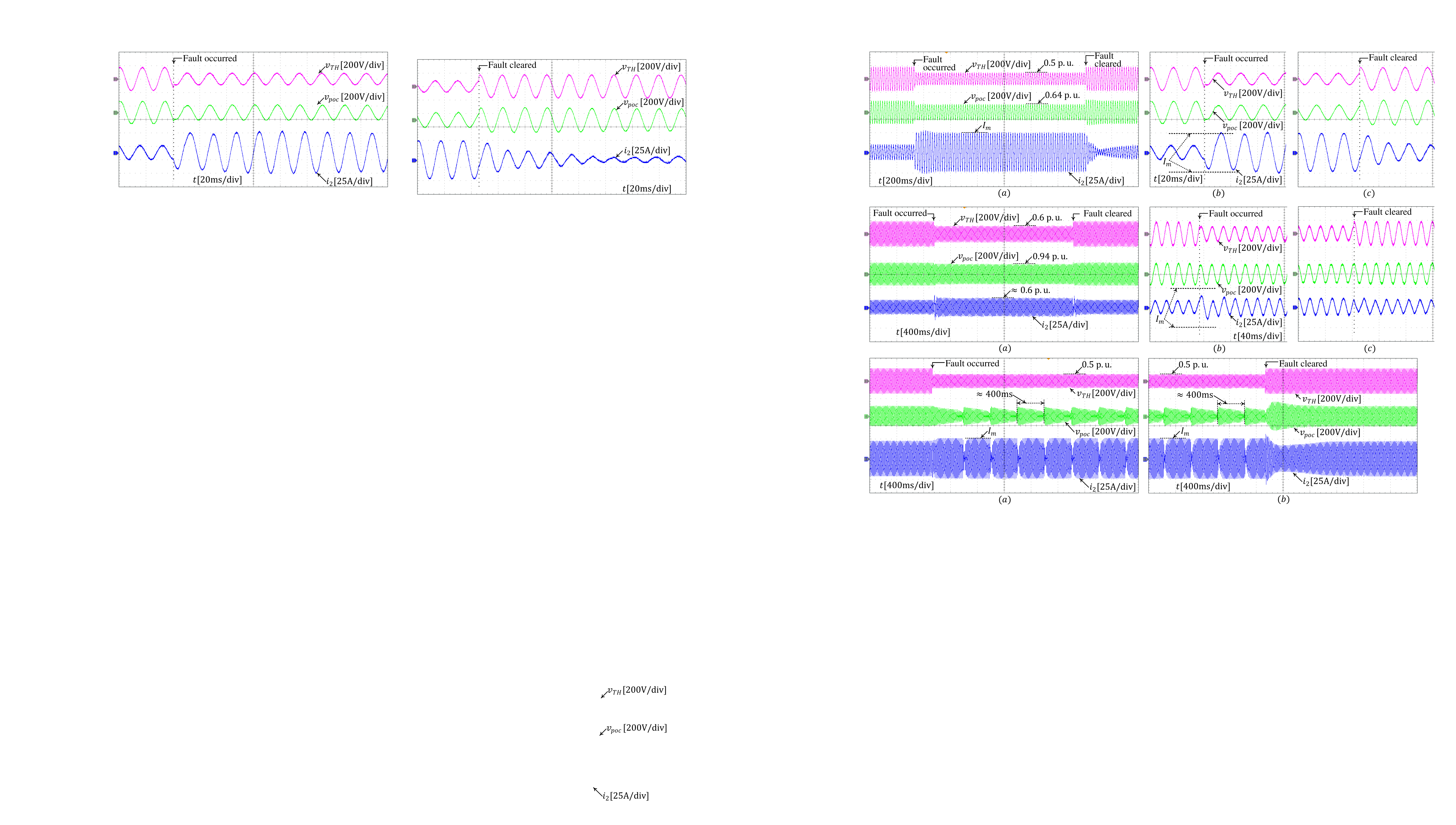}}
	\caption{Experimental results for \emph{Case II} - (a) converter response to an upstream grid fault when converter current reaches the hardware limit, (b) zoomed-in response when fault occurs, (c) zoomed-in response during fault clearing.}
	\label{fig:exp_CaseII}
\end{figure*}

The advantages of the proposed phase-plane analysis using $\Dot{x}-x$ surfaces over the conventional approach using $\Dot{\delta}-\delta$ and $V-\delta$ curves become evident in the analysis for \emph{Case III}. In the conventional method, $\Dot{\delta}-\delta$ curve shows existence of equilibrium points ($\Dot{\delta}=0$); however, the $V-\delta$ plot shows a limit-cycle behavior. Although, the conventional method indicates a limit-cycle behavior, it fails to explain why and how such oscillatory behavior occurs. Contrarily, the proposed $\Dot{x}-x$ surfaces immediately reveals that although $\Dot{\delta}=0$ and $\Dot{V}=0$ can be reached separately in the phase-plane, $(\Dot{\delta},\Dot{V})\equiv (0,0)$ cannot be reached which leads to the limit-cycle response. 




The first-order dynamics of uVOC has a large-signal stability advantage over second-order controllers such as droop control or VSM. Second order controllers are constrained by a critical clearing angle (CCA); the system loses stability if the fault is not cleared and/or power reference set-points are modified so as to ensure that a stable equilibrium exists, before the power angle exceeds the CCA. However, no such constraint exists for uVOC due to its first-order nature. To illustrate that uVOC is not constrained by a critical clearing angle, we chose multiple fault-clearing angles, i.e., points (temporally uniformly spaced) on the limit-cycle at which the fault is cleared by raising the grid voltage to $v_{TH}=1$ p.u.; the system trajectory is solved for each such initial condition. Evidently, irrespective of the fault clearing angle, all trajectories, shown in Fig.~\ref{fig:xDotxFLT_CaseIII}(c), reach the post-fault equilibrium point which is the same as the pre-fault equilibrium point $a_s$. The simulated converter response corresponding to \emph{Case III} is shown in Fig.~\ref{fig:xDotxFLT_CaseIII}(d). A limit-cycle behavior is observed during the fault. The simulated period, i.e., $\approx 393$ ms, of the limit-cycle behavior matches with that obtained from the trajectory analysis. Stable operation is retained once the fault is cleared. It is worth noting that the current-limiting controller effectively limits the current at $I_m$ even during the fault condition when no equilibrium can be reached.   

\section{Experimental Results}
The experiments are performed using a single-phase system with identical passive components and control parameters as those used for analysis and simulation of the three-phase system in the preceding sections; the power ratings are scaled to $33\%$ to obtain an equivalent system. The detail system parameters are listed in Appendix~\ref{apndx:simExpSetup}.   

\subsection{Case I}
The first set of experiments are performed for the system and grid conditions identical to those in Section \ref{sec:tsaCaseI}. The converter response is shown in Fig.~\ref{fig:exp_CaseI}(a). As predicted by the analysis and simulation, in response to the grid voltage sag to $v_{TH}=0.6$ p.u. the converter output current does not exceed the current limit $I_m$ (see Fig.~\ref{fig:exp_CaseI}(b)) and the reactive current injection raises the converter terminal voltage to $v_{poc}\approx 0.94$. Once the grid voltage returns to the nominal value, normal operation is retained. The transients when the fault occurs and during the fault-clearing are shown in Fig.~\ref{fig:exp_CaseI}(b) and \ref{fig:exp_CaseI}(c), respectively.

\subsection{Case II}
The next set of experiments are performed at identical conditions as described in Section~\ref{sec:tsaCaseII}. In response to the grid voltage sag to $v_{TH}=0.5$ p.u., an over-current fault is detected once the converter current exceeds $I_m$ (see Fig.~\ref{fig:exp_CaseII}(b)); the controller increases its reactive power output while clamping the output current at $I_m$ which raises the converter terminal voltage to $0.64$ p.u. Once the fault is cleared, the converter returns to normal operation (see Fig.~\ref{fig:exp_CaseII}(a) and (c)).

\subsection{Case III}
\begin{figure*}[tb]
	\makebox[\linewidth][c]{\includegraphics[angle = 0, clip, trim=0cm 0cm 0cm 0cm,  width=0.95\textwidth]{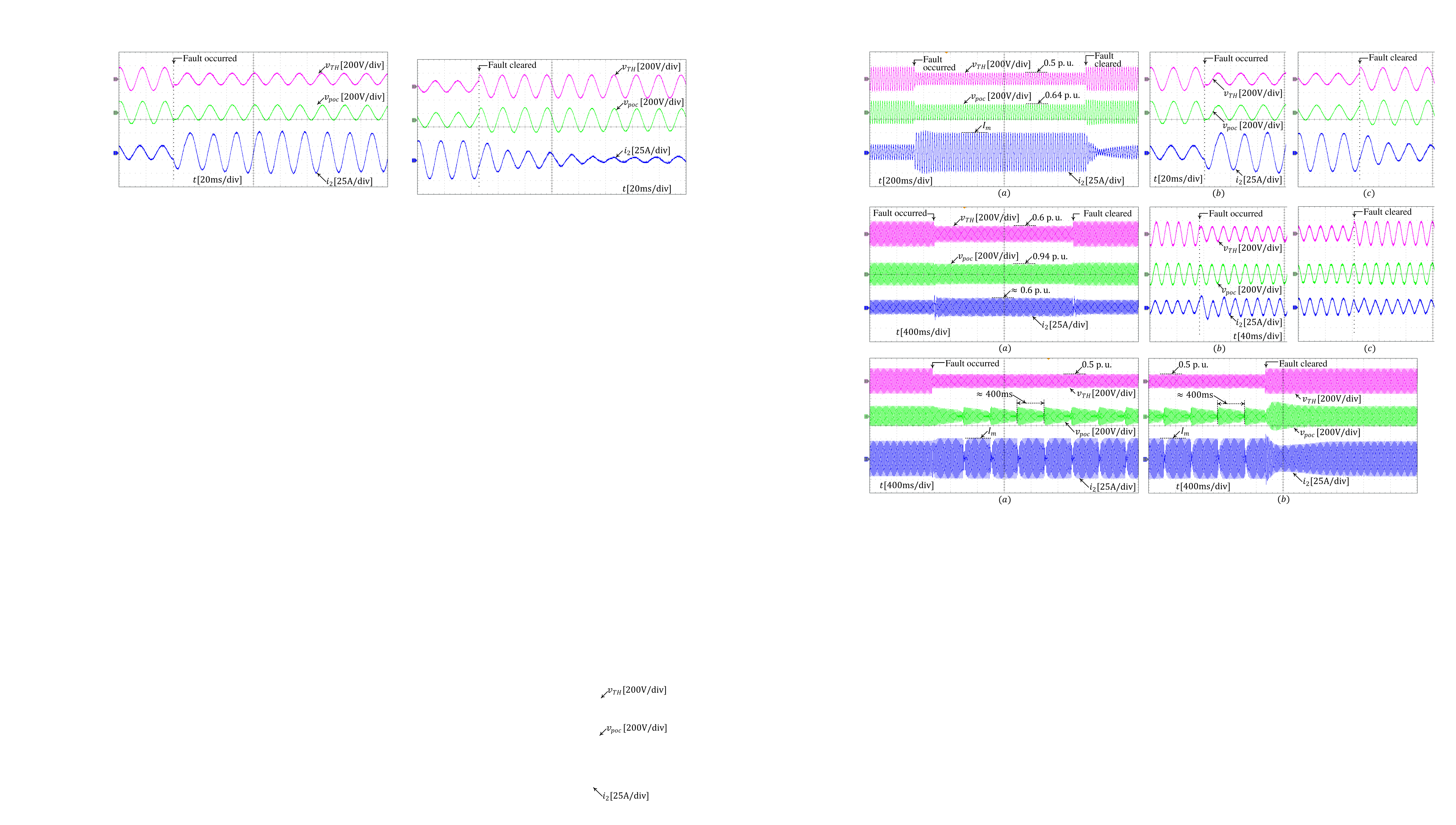}}
	\caption{Experimental results for \emph{Case III} - (a) converter enters a limit-cycle response during fault since no equilibrium point exists, (b) normal operation is retained as soon as the fault is cleared.}
	\label{fig:exp_CaseIII}
\end{figure*}

For the system and grid conditions described in Section~\ref{sec:tsaCaseIII}, the test results are shown in Fig.~\ref{fig:exp_CaseIII}. In response to the fault, the converter output current reaches $I_m$ and enters current saturated operation. As predicted by the analysis, a limit-cycle response with a period of $\approx 400$ms is observed (see Fig.~\ref{fig:exp_CaseIII}(a)). Note that the output current is limited to $I_m$ during the fault. Once the fault is cleared, the converter returns to normal operation quickly (see Fig.~\ref{fig:exp_CaseIII}(b)). 

\section{Conclusion}
A comprehensive modeling and analysis method has been developed for transient stability assessment of uVOC based grid-forming converters under both current constrained and unconstrained symmetrical faults. The proposed phase-plane analysis provides holistic insight into both voltage and power angle dynamics through a single graphical representation. Due to its first order nature, uVOC is not constrained by a critical clearing angle unlike second order droop based controllers; such property of uVOC is retained under current-constrained faults as well. The large signal nonlinear analysis demonstrates that uVOC with its fast response and built-in fault ride-through parameters exhibits strong stability characteristics. The modelling and analysis method has been validated through simulation and hardware experiments. 
\appendices
\section{Simulation and Experimental Setup}\label{apndx:simExpSetup}
The simulations and experiments are performed using a three-phase and an equivalent single phase VSC, respectively, with identical LCL filter and control parameters. The power ratings of the three-phase system is chosen as $\times 3$ of those for the single-phase system. The system parameters are listed in Table~\ref{TB:vscParam}. Note that two interleaved sub-phases are used in each phase and hence an effective switching frequency of $75$kHz is achieved which reduces the LCL filter requirements\cite{ecceGFMmpc}. The control parameters are listed in Table~\ref{TB:conParam}. All parameters are listed using $P_{rated}$ and $V_0$ as base quantities.

\renewcommand{\arraystretch}{1}
\begin{table}[htb]
\centering
\caption{Voltage Source Converter Ratings}
\begin{tabular}{ p{0.9cm}p{3.05cm}p{3.25cm}}
\hline
\hline
Symbol & Parameter & For $3\phi$ (for $1\phi$)\\
\hline
$S_{rated}$ & Rated power & $9\ \text{kVA}$ ($3\ \text{kVA}$)\\
$P_{rated}$ & Rated real power & $7.5\ \text{kW}$ ($2.5\ \text{kW}$)\\
$Q_{rated}$ & Rated reactive power & $5\ \text{kVAR}$ ($1.67\ \text{kVAR}$)\\
$V_0$ & Nom. (L-N RMS) voltage & $120\ \text{V}$\\
$\omega_0$ & Nom. frequency & $2\pi(60)\ \text{rad/s}$\\
$f_{sw}$ & Switching frequency & $37.5$ kHz\\
$f_s$ & Sampling frequency & $37.5$ kHz\\
$L_a$ & Converter-side inductor & $\approx 0.04\ \text{pu} \times 2$ interleaved\\
$L_g$ & Network-side inductor & $\approx 0.005\ \text{pu}$\\
$C_f$ & Filter capacitor & $\approx 0.004\ \text{pu}$\\
\hline
\hline
\end{tabular}
\label{TB:vscParam}
\vspace{0pt}
\end{table}

\renewcommand{\arraystretch}{1}
\begin{table}[htb]
\centering
\caption{Controller Parameters}
\begin{tabular}{ll|ll}
\hline
\hline
$\eta$ & $19.95$ & $\mu$ & $7.1\times 10^{-4}$\\
$\tau_f$ & $0.11$ & $R_0$ & $0.43$ p.u.\\
$L_{v0}$ & $0.29$ p.u. & $R_{v0}$ & $0.04$ p.u.\\
$\omega_{b}$ & $2\pi(600)$ rad/s & $I_m$ & $1.2$ p.u.\\
\hline
\hline
\end{tabular}
\label{TB:conParam}
\vspace{0pt}
\end{table}

\bibliographystyle{IEEEtran}
\bibliography{UOC}

\end{document}